\documentclass[journal]{IEEEtran}
\usepackage{graphicx}
\usepackage{dcolumn}
\usepackage{bm}
\usepackage{amssymb}
\usepackage{amsmath}
\usepackage{subfigure}
\usepackage{physics}

\begin{document}

\title{Enhancement of Thermal Spin Transfer Torque via Bandpass Energy Filtering}

\author{Pankaj Priyadarshi, Abhishek Sharma and Bhaskaran~Muralidharan
	
	\thanks{The authors are with the Department of Electrical Engineering, IIT Bombay,
		Mumbai 400076, India (e-mail: bm@ee.iitb.ac.in, priyadarshi@ee.iitb.ac.in).}
}

\maketitle

\date{\today}

\begin{abstract}
We propose employing the energy bandpass filtering approach in a magnetic tunnel junction device as a route to enhance the thermal spin transfer torque. Using the spin-resolved non-equilibrium Green’s function formalism, we propose to harness the optical analog of anti-reflective coatings in a heterostructure MTJ device, and report a large boost in spin torque in the linear regime of temperature bias. In particular, we discuss the position of transmission function with respect to the Fermi energy that caters the maximum thermal effect. A ``boxcar" transmission feature of resulting from the anti-reflective configuration enhances charge and spin transport through the structure in comparison to a normal superlattice configuration. The thermally excited spin transfer torque is enhanced by almost five times with our device designs. Although, the thermally driven spin torque is much smaller than the voltage driven torque, this technique provides an energy efficient way to switch the magnetization. This opens up a new viable area for spintronics applications and with the existing advanced thin-film growth technology, the optimized superlattice configurations can be achieved.
\end{abstract}


\section{Introduction}
Magnetization manipulation is one of the most important issues in spintronics applications. Customarily, magnetization switching is achieved by the use of an external magnetic field or by the current induced spin transfer torque (STT) \cite{Slonczewski1989, Slonczewski1996, Berger1996, Ralph2008}, though there is a fundamental difference between these two approaches. Spin-polarized current, traversing a fixed ferromagnetic (FM) layer through a spin valve structure exerts a torque on the magnetic moment of the free ferromagnetic layer, due to the interaction between the conduction electron spin and the local magnetization, which is known to reduce the switching energy consumption. Thereby, the STT effect is a promising technology for applications in torque driven magnetization switching, magnetic sensors \cite{Abhishek2016}, domain wall motion, high-density non-volatile memory devices such as STT-MRAM (magnetic random access memories) \cite{Parkin1999, Locatelli2014}, STT-Ddiodes \cite{Tulapurkar2005}, and STT-nano-oscillators (STT-NO)  \cite{Abhishek2017}.

However, explorations on new and reliable mechanisms that can switch the magnetization more efficiently are still being carried out. A temperature gradient can also manipulate the magnetization of free/soft FM layer in structures like magnetic tunnel junctions (MTJs) \cite{Hatami2007, Slonczewski2010, Jia2011, Heiliger2014}, termed as thermal spin transfer torque (TSTT). The TSTT phenomenon relies only on the temperature differences across the device, in contrast to the traditional voltage bias dependent STT, and thus potentially provides an efficient way for magnetization switching. The coupling of thermal effects in addition to the electrical and magnetic properties of the nanostructure has given rise to the new and active field of ``spin caloritronics" \cite{Gravier2006, Uchida2008, Adachi2013}. Thus, theoretical studies have opened up a variety of novel effects, including thermally excited spin currents \cite{Oleksandr2006}, spin Seebeck effect \cite{Walter2011, Adachi2013}, nanomagnet heat pumps or power generators \cite{Arrachea2015}, and devices such as rotational nanomotors, etc., to name a few.

The most common structure for an MTJ is a trilayer structure, which consists of an insulating barrier between two ferromagnetic layers- a fixed layer and a free layer. The magnetic properties of spin devices such as the tunnel magneto-resistance (TMR) and voltage-dependent STT effect are explored extensively in MTJs. Progressively, it is shown that the voltage dependent STT may be greatly improved by using various configurations of multiple barrier MTJs \cite{Niladri2015, Abhishek2016}. Thus, some groups have a keen interest in investigating the multi-barrier structures (i.e., superlattice) with the intent of further improvement in the desired benchmarks. Although, the superlattice (SL) structures have been well studied in the field of photonics, electronics, and thermoelectrics \cite{Pankaj2018, Swarnadip2018}, there is a dearth in understanding SL structures for spintronic applications. The incorporation of a multi-layer structure in MTJ which comprises of alternate layers of an insulator and nonmagnetic metal (NM) sandwiched between the two FMs, in particular, can enhance the TMR \cite{Chen2014, Abhishek2018}. The use of SL as a functional region leads to the formation of passbands and forbidden bands in energy for the coherent electron transport \cite{Pacher2001, Morozov2002}. This bandpass feature of the transmission profile can be optimally utilized to improve the flow of charge and spin currents. A very high TMR of more than $10^4 \%$ in a bandpass SL \cite{Abhishek2018} has been reported. The TSTT effect in a double barrier (i.e., resonant tunneling) MTJ has also been studied \cite{Chen2017-1}, which shows an advantage over a single barrier MTJ.  In this study, we propose the use of an anti-reflective region in the SL structure in the context of TSTT. This additional region, when appended across the normal SL structure, enables the carriers to meet almost zero reflection in the passband energy range compared to the normal SL. 

In particular, we discuss the position of transmission function with respect to the Fermi energy that caters the maximum thermal effect. A ``boxcar" transmission feature of resulting from the anti-reflective configuration enhances charge and spin transport through the structure in comparison to a normal superlattice configuration. The thermally excited spin transfer torque is enhanced by almost five times with our device designs. Although, the thermally driven spin torque is much smaller than the voltage driven torque, this technique provides an energy efficient way to switch the magnetization.

The paper is organized as follows. In section II, we describe the SL configuration and the simulation setup for numerical calculations of spin and charge transport model. The simulation results are presented and discussed in section II. We conclude the article in section IV.

\section{Device Structure and Simulation Formalism}
\begin{figure}
	\centering
	\includegraphics[height=0.25\textwidth,width=0.51\textwidth]{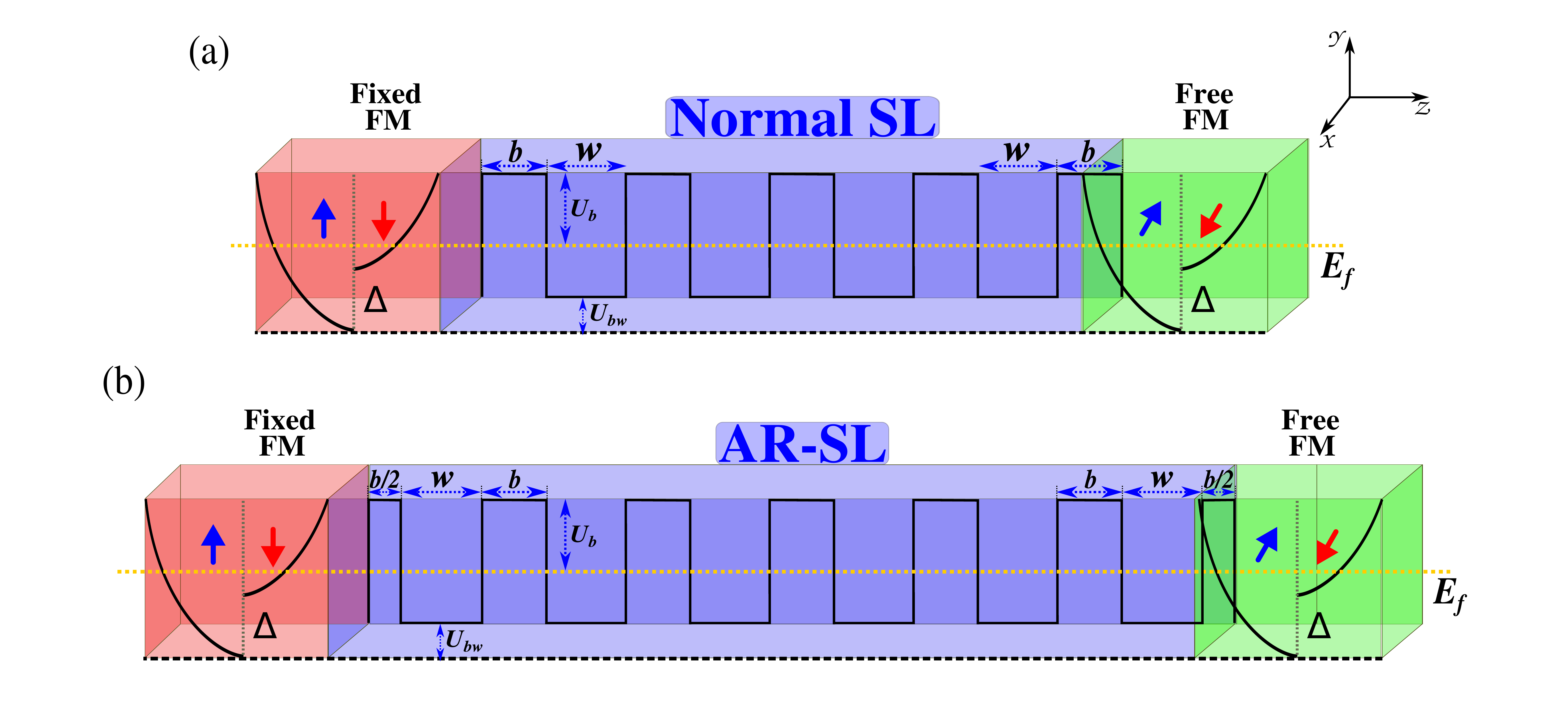}
	\caption{Device schematics: A typical MTJ with superlattice-like structure in the central region sandwiched between two FMs which is well represented by the Stoner model of band splitting. (a) NSL is the regular SL structure having a constant well width and barrier thickness along the transport $(\hat{z})$ direction. (b) AR-SL is the anti-reflection enabled SL, in which two additional barriers of half the thickness of a regular barrier are attached after a well width at both the ends.}
	\label{Device}
\end{figure}

Figure~\ref{Device} shows the device schematic of a typical SL-MTJ structure, depicted in the three different regions. The central region consists of a multi-barrier superlattice, sandwiched between two FM contacts. The red shaded part is a fixed FM maintained at a high temperature in comparison to the green shaded free FM. To incorporate the physics of spin transport, we use a simplified free-electron Stoner model of ferromagnetism. This assumes that the minimum of the parabolic conduction bands for spin-up and spin-down electrons have a relative shift in energy due to an exchange interaction, denoted by an amount of spin-splitting $\Delta$ in Fig.~\ref{Device}. The blue shaded region is a SL structure made from an alternate layer of an insulating barrier and a nonmagnetic metal whose lower edge of the conduction band differs by an amount of $U_{bw}$ with the conduction band edge of the FMs. Here, $w$ is the well width, $b$ is the barrier thickness, and $U_b$ is the barrier height of the SL configuration above the equilibrium Fermi energy $E_f$ (sketched in yellow dash line).

Our focus is on the central part of the device that effectively gives rise to a probability of electron transmission. The configuration in Fig.~\ref{Device}(a) depicts a regular well-barrier structure in series along the transport $(\hat{z})$ direction, labeled as \textit{Normal SL} (NSL). Likewise, Fig.~\ref{Device}(b) is a normal SL, but sandwiched between two barriers of half the thickness ($b/2$) of the regular barriers that serves as the anti-reflective (AR) region \cite{Pacher2001}, labeled as \textit{AR-SL}. The inclusion of AR-SL in the SL-MTJ structure leads to the formation of an energy bandpass transmission lineshape across it. Energy bandpass filtering can be attained via many other structural configurations such as Gaussian barrier thicknesses or heights \cite{Pankaj2018}, but at the cost of simulation time.

The prescription of quantum transport is given by the spin-resolved non-equilibrium Green's function (NEGF) formalism. The device set up is described by the varying effective-mass Hamiltonian $[H]$ using the nearest neighbor tight-binding model \cite{QTDatta}. We add a potential energy term $[U]$ to the matrix $[H]$, that takes care of the structural variation in the device region. We employ a one band spin-dependent NEGF formulation for the transmission and currents calculation in the device. To start with, the energy-resolved retarded Green’s function matrix $[G(E)]$ evaluated from the device Hamiltonian matrix $[H]$ is given by

\begin{equation}
[G(E)]=[E\mathbb{I}-H-U-\Sigma_H-\Sigma_C]^{-1},
\label{eqG}
\end{equation}

where $\mathbb{I}$ is the identity matrix. The quantities $\Sigma_H$ and $\Sigma_C$ are the self-energy matrices of the hot and cold ferromagnet contacts. The charge and spin currents through the device along the transport direction ($\hat z$) are given by 

\begin{equation}
I=q\int dE \; Real[Trace(\hat{I}_{op})],
\label{eqIc}
\end{equation}

and

\begin{equation}
I_s=q\int dE \; Real[Trace(\hat{I}_{op}. \sigma_s)]
\label{eqIs}
\end{equation}

respectively, where $q$ being the electronic charge, $\sigma_s$ is the Pauli matrices, and $\hat{I}_{op}$ is a current operator matrix of ($2\times2$) order in spin space, obtained through two adjacent lattice points $j$ and $j+1$, given by \cite{MesoDatta}

\begin{equation}
I_{op}(j,j+1) = \frac{i}{\hbar}\Big(H_{j,j+1}G_{j+1,j}^n - G_{j,j+1}^nH_{j+1,j}\Big),
\label{eqIop}
\end{equation}

where $\hbar$ is the reduced Planck's constant, $H$ being the Hamiltonian matrix of the system and $G^n$ is a diagonal element of the energy-resolved electron correlation matrix $[G^n(E)]$, given by

\begin{equation}
[G^n(E)]=[G][\Gamma_{H} f_H + \Gamma_{C} f_C][G]^{\dagger}.
\label{eqGn}
\end{equation}

Here, $\Gamma_H$ and $\Gamma_C$ are the spin-dependent broadening matrices of hot and cold FM contacts. The occupation probability of the hot and the cold contacts are respectively given by the Fermi$-$Dirac distribution function $f_H(E,\mu_H,T_H)$ and $f_C(E,\mu_C,T_C)$.

The spin current, in general, is a rate of flow of spin angular momentum. This momentum can act as a torque on the local magnetization of the free FM \cite{Ralph2008}. Further, we incorporate the Landau-Lifshitz-Gilbert-Slonczewski (LLGS) equation \cite{Slonczewski1996} that describes the magnetization dynamics of the free FM in the presence of spin current:

\begin{equation}
(1+\alpha^2)\frac{\delta\hat{m}}{\delta t} = -\gamma\hat{m} \times \hat{M} - \gamma\alpha \Big(\hat{m} \times \hat{M} \Big) - \vec{\tau}_{spin},
\label{eqLLGS}
\end{equation}

where $\hat{m}$ ($\hat{M}$) is the unit vector along the direction of magnetization of the free (fixed) FM, $\gamma$ is the gyromagnetic ratio of the electron, $\alpha$ is the Gilbert-damping parameter. The vector quantity $\vec{\tau}_{spin}$ in Eq.~(\ref{eqLLGS})  is the spin torque exerted on the free FM layer, given by

\begin{equation}
\vec{\tau}_{spin} = \frac{\gamma \hbar}{2qM_sV} \Big[\Big(\hat{m} \times (\hat{m} \times \vec I_s)\Big) - \alpha (\hat{m} \times \vec I_s) \Big].
\label{eqTorque}
\end{equation}

Here, $M_s$ and $V$ respectively are the saturation magnetization and the volume of free FM layer. Furthermore, we resolve the spin current $\vec{I}_s$ as

\begin{equation}
\vec{I}_s = I_{s,m} \hat{m} + I_{s,\parallel} \hat{M} + I_{s,\perp} \hat{M} \times \hat{m}.
\label{eqIspin}
\end{equation}

From Eqs. (\ref{eqLLGS}), (\ref{eqTorque}) and (\ref{eqIspin}), the magnetization dynamics is solved under the steady state condition. We now break the spin transfer torque into two components: one in-plane (formed by $\hat x- \hat y$ plane)

\begin{equation}
I_{s,\parallel} = \frac{\vec{I}_s . \Big((\hat{m} \times \hat{M}) \times \hat{m}\Big)}{1-(\hat{M} . \hat{m})^2},
\label{eqIsPar}
\end{equation}

and one out-of-plane component (formed by $\hat y- \hat z$ plane)

\begin{equation}
I_{s,\perp} = \frac{\vec{I}_s . (\hat{M} \times \hat{m})}{1-(\hat{M} . \hat{m})^2}.
\label{eqIsPer}
\end{equation}

The parallel (in-plane) component of the torque is clearly responsible for the magnetization switching. However, the role of the perpendicular (out-of-plane) torque is not much clear theoretically. Therefore, we focus our analysis only on the parallel component of the spin transfer torque (STT) and thus, we can refer STT with the term $I_{s||}$ interchangeably. Since the Fermi function depends on the temperature $T_{H(C)}$ of the contacts, that leads to open the window for current to flow. Hence, the torque exerted on the magnetization of the free magnet is termed as thermal spin transfer torque (TSTT).

\section{Results and Analysis}

\begin{figure}
	\subfigure[]{\includegraphics[height=0.18\textwidth,width=0.225\textwidth]{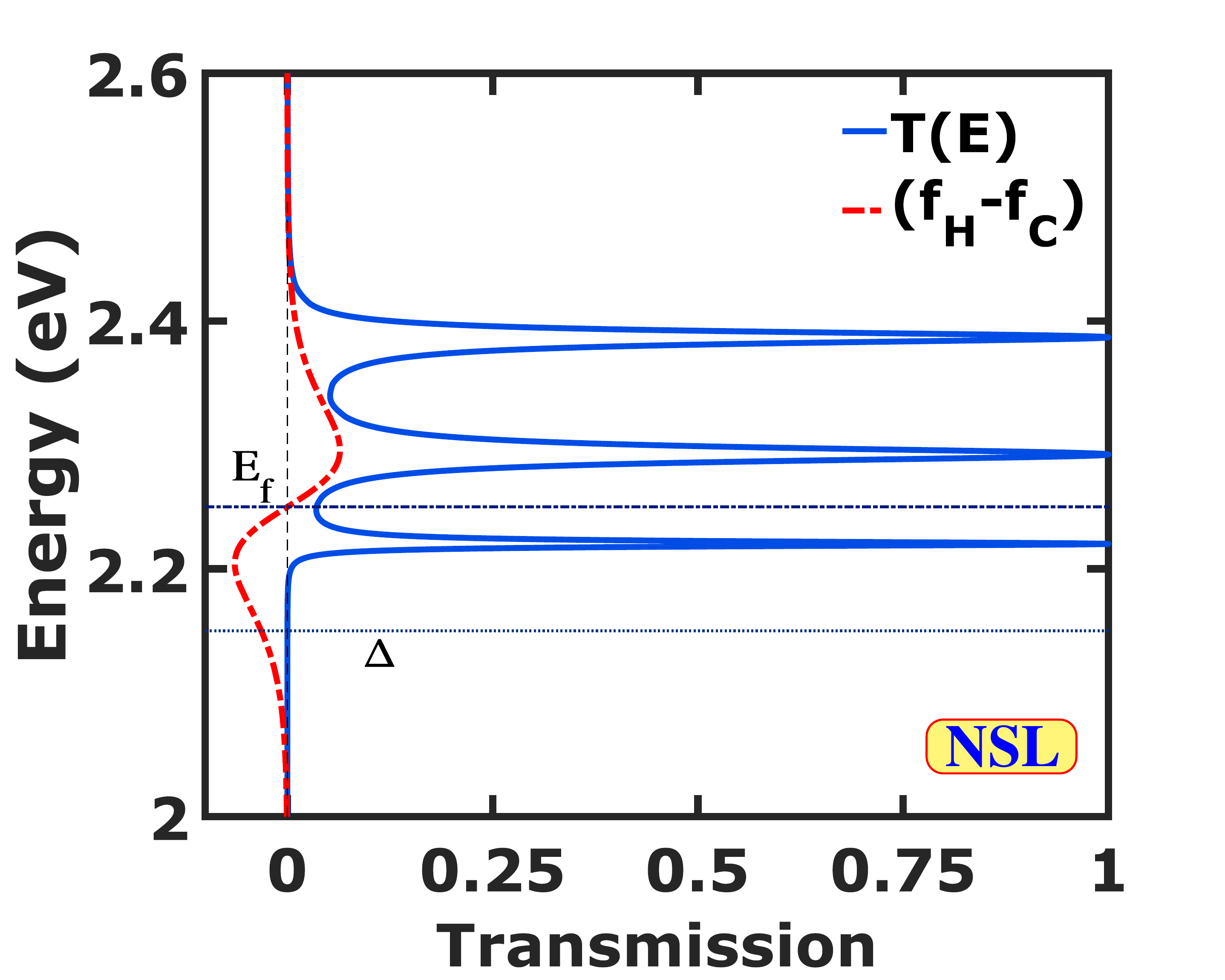}\label{NSL_4B_TM_f1-f2_EP}}
	\quad
	\subfigure[]{\includegraphics[height=0.18\textwidth,width=0.225\textwidth]{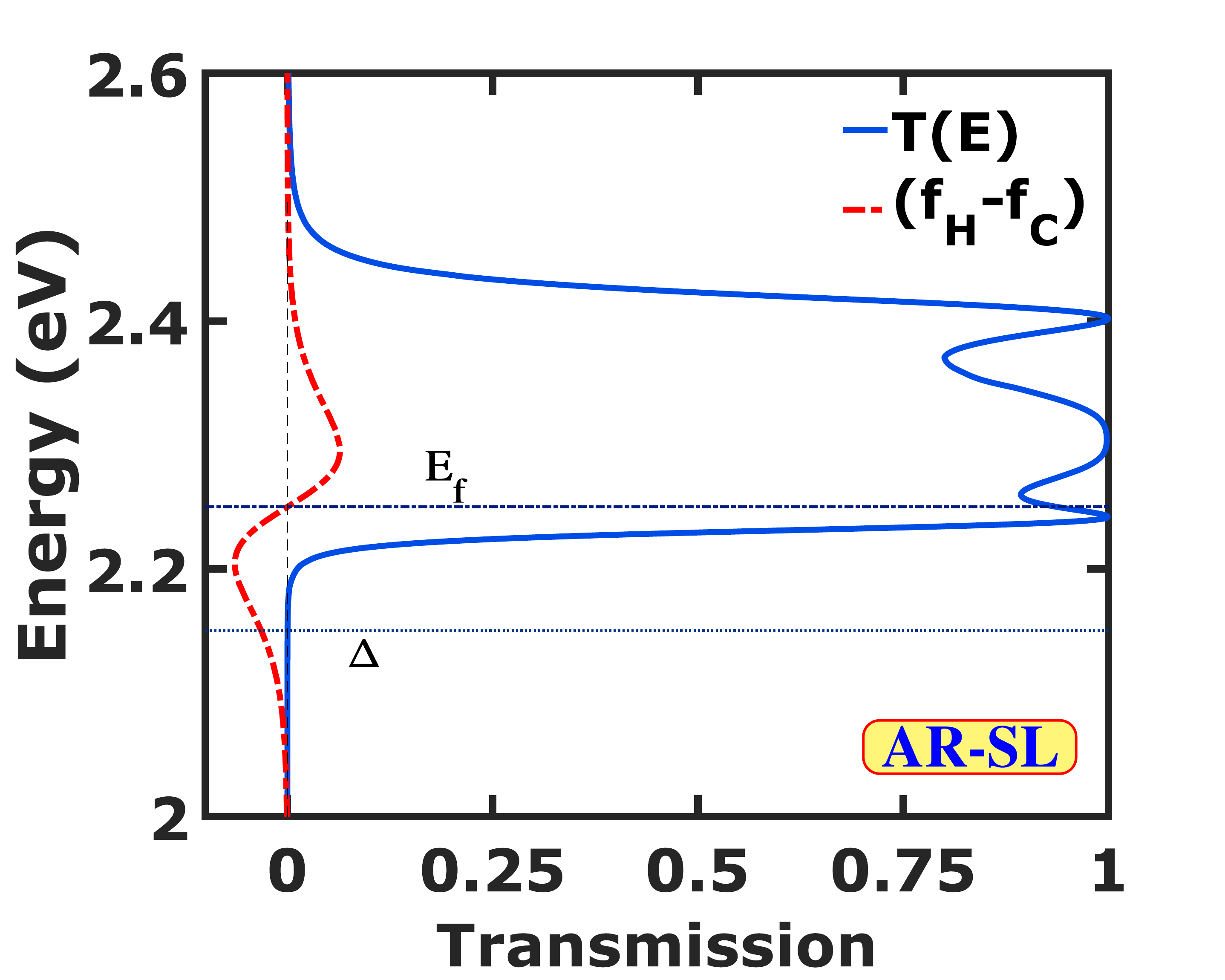}\label{ARSL_4B_TM_f1-f2_EP}}
	
    \caption{Color online: (a) Transmission coefficient as a function of energy $T(E)$ through a 4-barrier NSL (solid blue curve) for the first transverse mode with the Fermi function difference $(f_H-f_C)$ (red dash-dot line) at a temperature difference of 10 K between the contacts. The labels $E_f$ and $\Delta$ indicate the Fermi level and exchange splitting, respectively. Similarly (b), the transmission coefficient as a function of energy $T(E)$ through a 4-barrier AR-SL (solid blue curve) with Fermi function difference (red dash-dot line) at a temperature difference of 10 K between the contacts.  Plots are zoomed in only for the first miniband.}
	\label{4B_TM_f1-f2_EP}
\end{figure}

In the following, we discuss the simulation results of different SL configurations of MTJ family in the context of thermal spin transfer torque. We use CoFeB as the FM contacts with its Fermi energy $E_f=2.25~eV$ and exchange splitting energy $\Delta=2.15~eV$. The effective masses of electron in the FMs, the barrier (MgO), and the well (NM) regions are $m_{FM}=0.8m_e$, $m_{b}=0.18m_e$, and $m_{w}=0.9m_e$, respectively \cite{Datta2012}, with $m_e$ being the free electron mass. The conduction band offset between the barrier-well interface is $U_b=0.76~eV$ above the Fermi energy \cite{Kubota2008}. We added up each current through all the transverse modes that fit into the cross-sectional area of $50 \times 50~nm^2$ nanowire. These transverse modes of electron conduction may be visualized as a set of parallel sub-band diagrams with an offset. In the process of calculating the TSTT, it is assumed that the temperature difference between the contacts are so small that the linear response theory can be applied.

\subsection{Thermal bias and average electrochemical potential}
\begin{figure}
	\subfigure[]{\includegraphics[height=0.18\textwidth,width=0.225\textwidth]{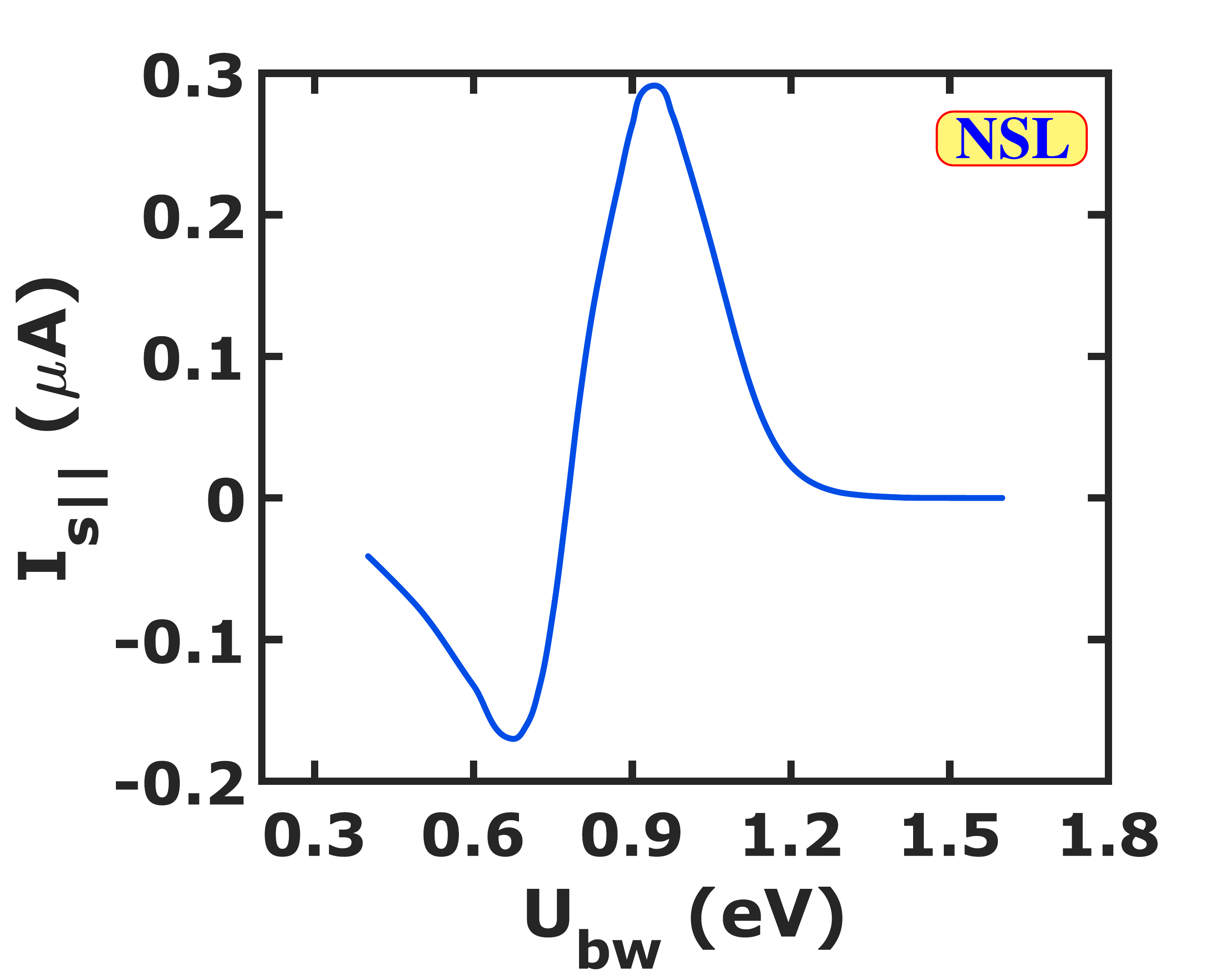}\label{NSL_4B_IsPara_Ubw_90}}
	\quad
	\subfigure[]{\includegraphics[height=0.18\textwidth,width=0.225\textwidth]{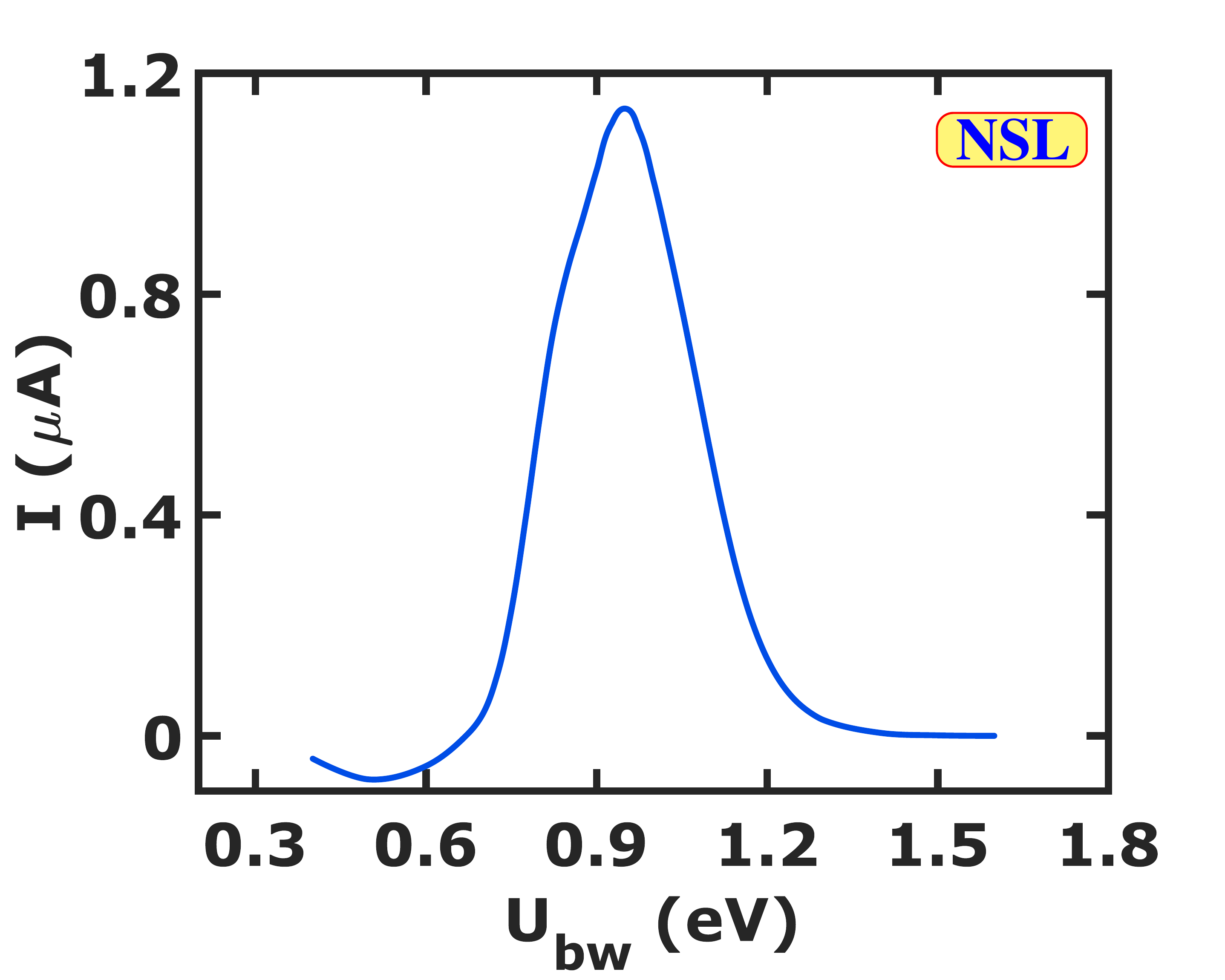}\label{NSL_4B_Ic_Ubw_90}}
	\quad
	\subfigure[]{\includegraphics[height=0.18\textwidth,width=0.225\textwidth]{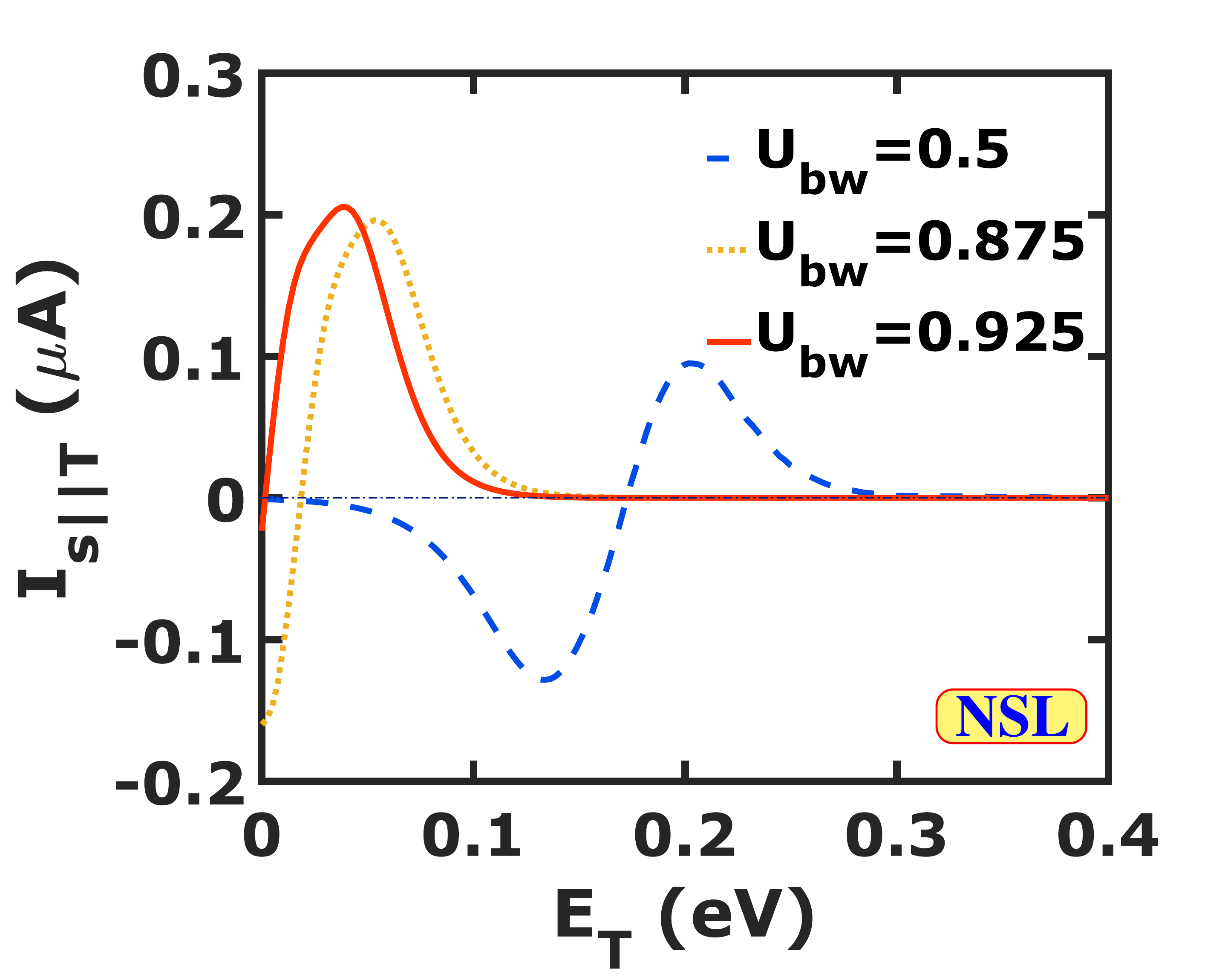}\label{NSL_4B_IsPara_linMode_Ubw_90}}
	\quad
	\subfigure[]{\includegraphics[height=0.18\textwidth,width=0.225\textwidth]{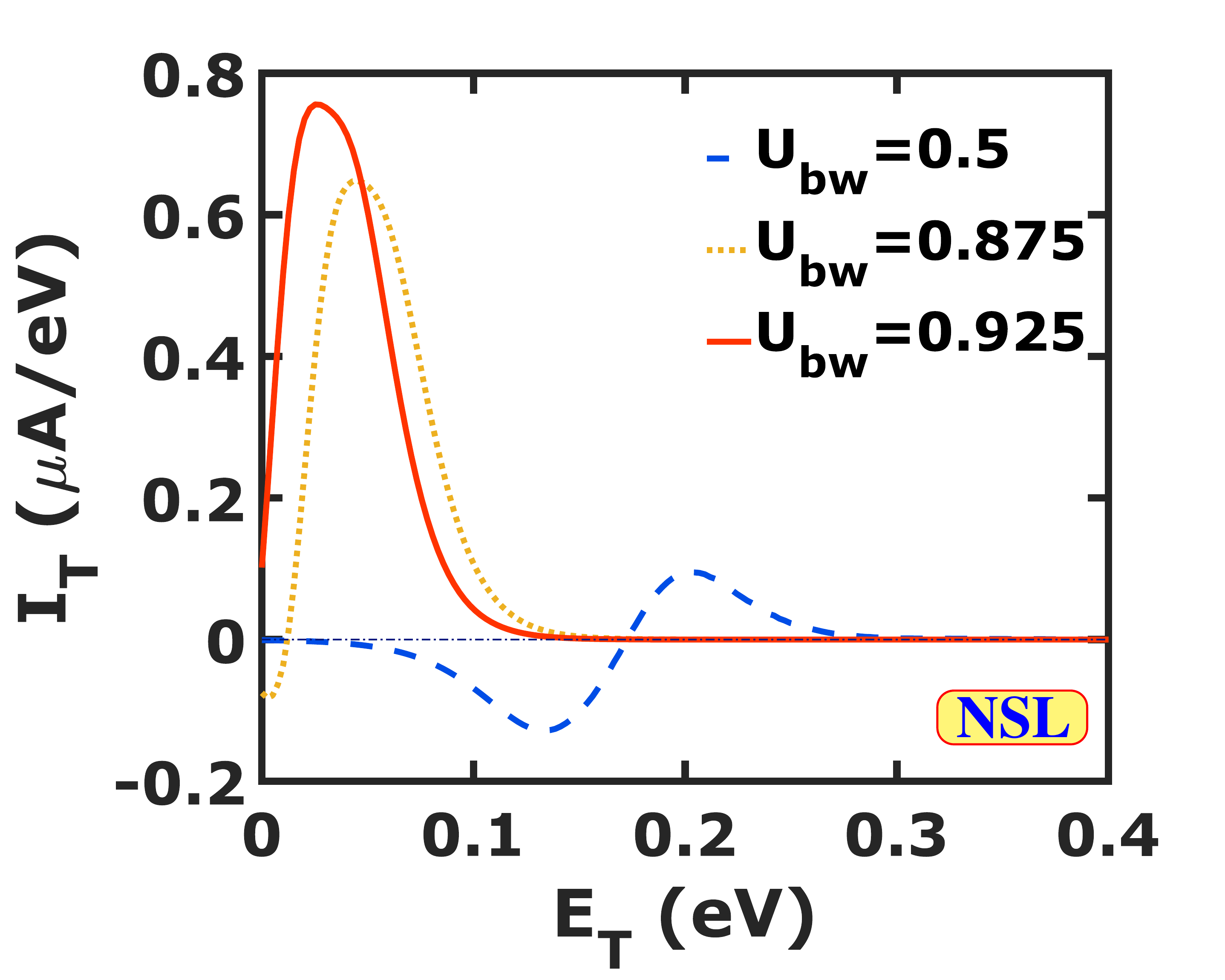}\label{NSL_4B_Ic_linMode_Ubw_90}}
	
	\caption{(a) Parallel component of thermal spin transfer torque ($I_{s||}$) and (b) charge current ($I$) as a function of $U_{bw}$ for NSL in the perpendicular configuration of the fixed and free FMs, while keeping $V=0$ and $\Delta T=10 K$. Transverse mode profile for (c) parallel spin current and (d) charge current in a 4-barrier NSL.}
	\label{NSL_4B_I_Upw_90}
\end{figure}

As mentioned above, it is worth to discuss only the parallel component of the TSTT acting on the free FM layer. It is well established that to get the advantage of thermoelectric effect, the transmission function must be just above the Fermi energy $E_f$ \cite{MesoDatta}. To illustrate this fact, we plot, in Fig.~\ref{4B_TM_f1-f2_EP}, the transmission probability of first transverse mode with the Fermi function difference between the hot and cold contacts, ($f_H-f_C$) as a function of energy, due to temperatures $T_H$ \& $T_C$ at the contacts, respectively. In this case, $(f_H-f_C)$ is anti-symmetric with respect to $E_f$. It is noted that the magnitude of current will be greater when $T(E)$ encloses a maximum area above $E_f$. Thus, we can say that the thermal bias effect is more sensitive than the voltage bias effect, that is, in general, symmetric about $E_f$ and rectangular in shape with unity probability. From Fig.~\ref{ARSL_4B_TM_f1-f2_EP}, we note that the transmissivity (area under $T(E)$ curve) in the AR-SL configuration is maximum in comparison with the NSL configuration. This feature of AR enabled structure can be understood by the two basic key concepts: 1) the AR layer should be a Bragg reflector in energy and 2) the potential profile of the AR layer should be such that the electronic state at that energy becomes a Bloch eigenstate of the central region \cite{Martorell2004}.

In order to do so in our device configurations, we vary the parameter $U_{bw}$ in such a way that we get the maximum thermal currents. Here $U_{bw}$ refers to the well height as shown in Fig.~\ref{Device} \cite{Niladri2015}. Figure~\ref{NSL_4B_IsPara_Ubw_90} shows the variation of parallel components of TSTT ($I_{s||}$) as a function of $U_{bw}$, in response to $\Delta T$ only. It shows a sinusoidal behavior that reflects the anti-symmetric nature of the Fermi function difference due to temperature \cite{Walter2011}. For 4-barrier NSL the maximum $I_{s||}$ is found at $U_{bw}=0.925~eV$ for $\Delta T=~10K$. At this value of the well height, the transmission captures only the positive lobe of the Fermi function difference. It may be understood by the transverse mode spin current profile, that sum up to get the complete current through the device, as shown in Fig.~\ref{NSL_4B_IsPara_linMode_Ubw_90}. Here, we note that as the value of $U_{bw}$ increases, the negative lobe of the current gets eliminated and also increases the total $I_{s||}$, thus enabling us to capture the full positive swing due to the thermal bias. Similarly, the charge current $I$ reaches a maximum at the same $U_{bw}$, as shown in Fig.~\ref{NSL_4B_Ic_Ubw_90}. The plot of transverse mode charge current profile is depicted in Fig.~\ref{NSL_4B_Ic_linMode_Ubw_90}.

\begin{figure}
	\subfigure[]{\includegraphics[height=0.18\textwidth,width=0.225\textwidth]{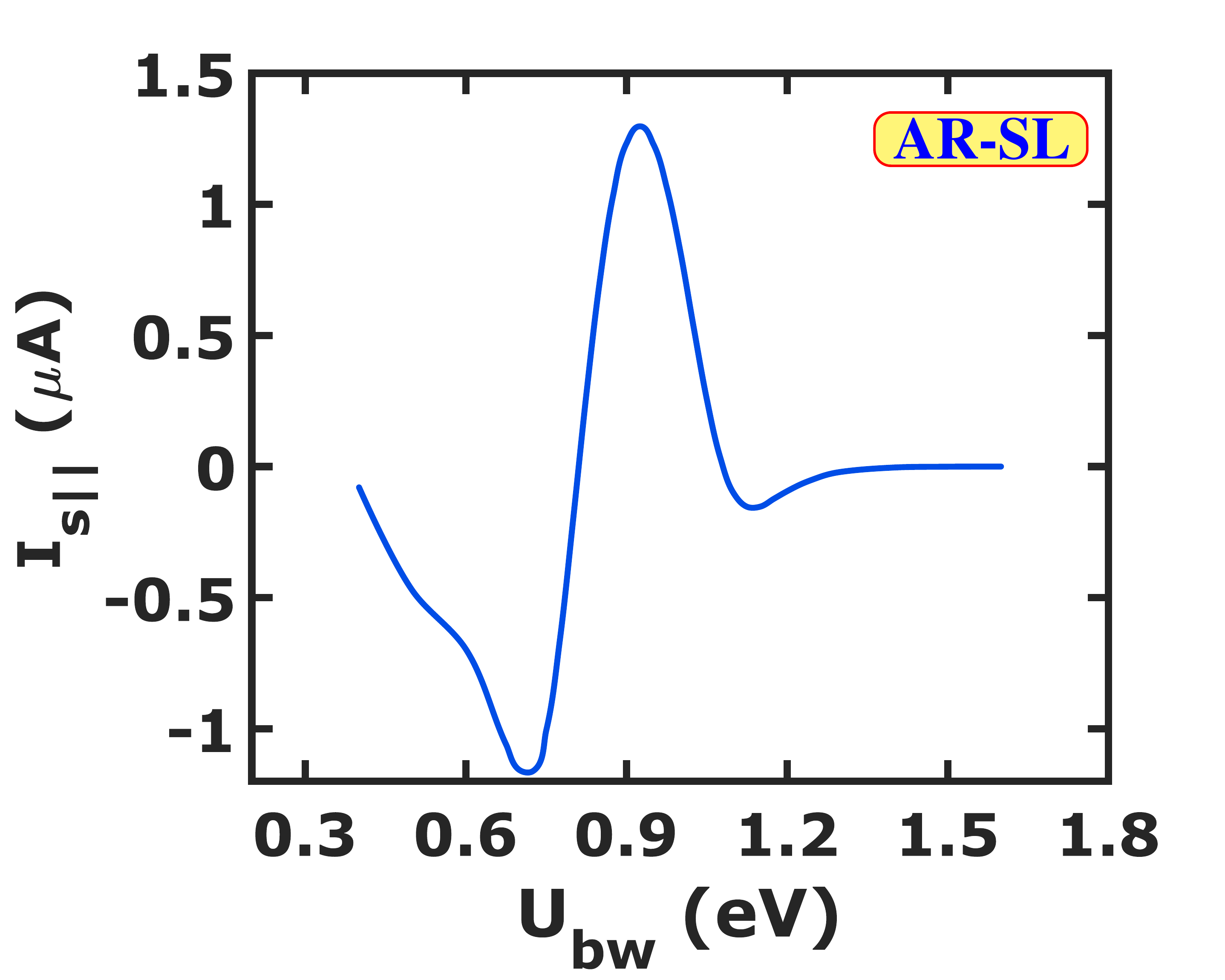}\label{ARSL_4B_IsPara_Ubw_90}}
	\quad
	\subfigure[]{\includegraphics[height=0.18\textwidth,width=0.225\textwidth]{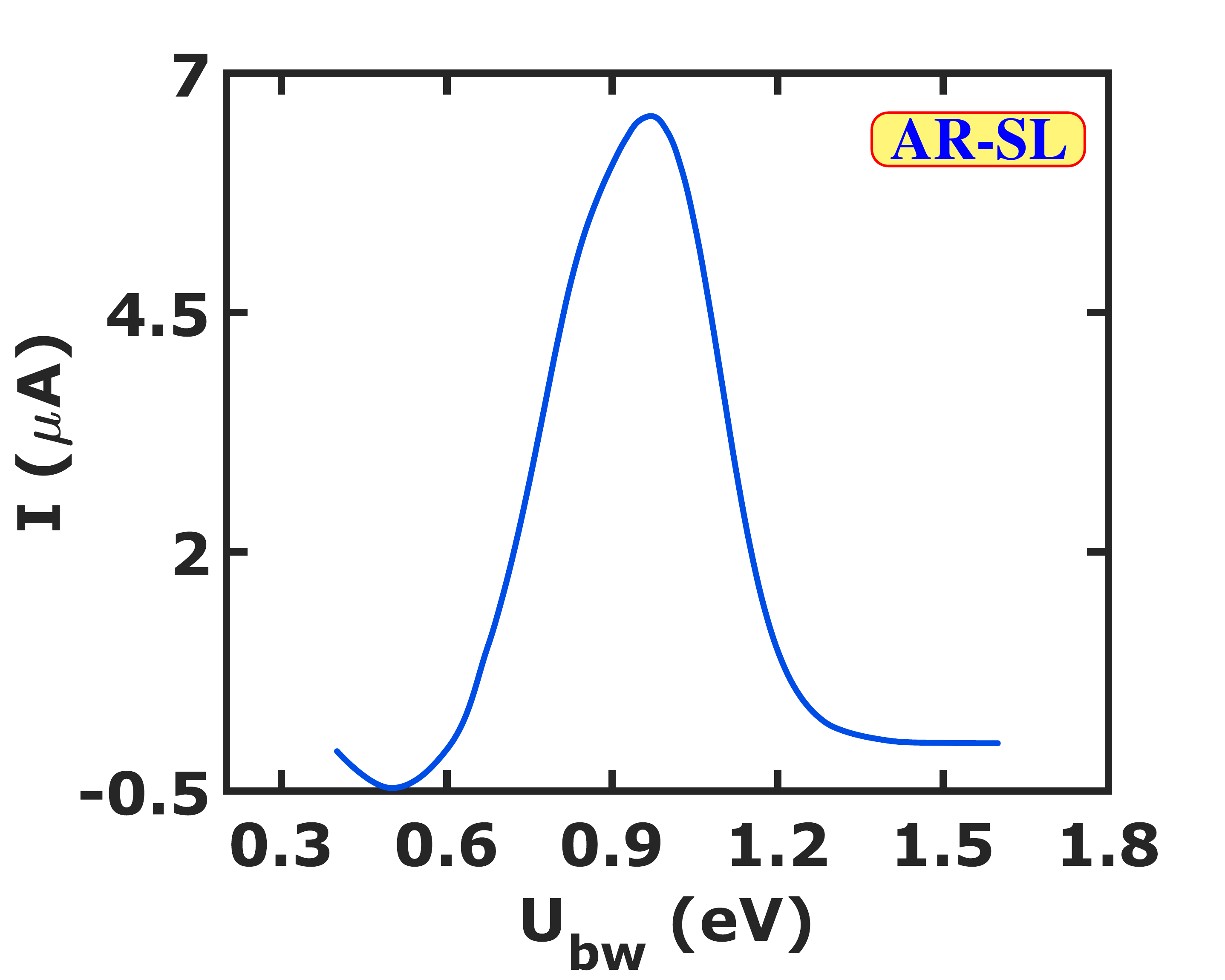}\label{ARSL_4B_Ic_Ubw_90}}
	\quad
	\subfigure[]{\includegraphics[height=0.18\textwidth,width=0.225\textwidth]{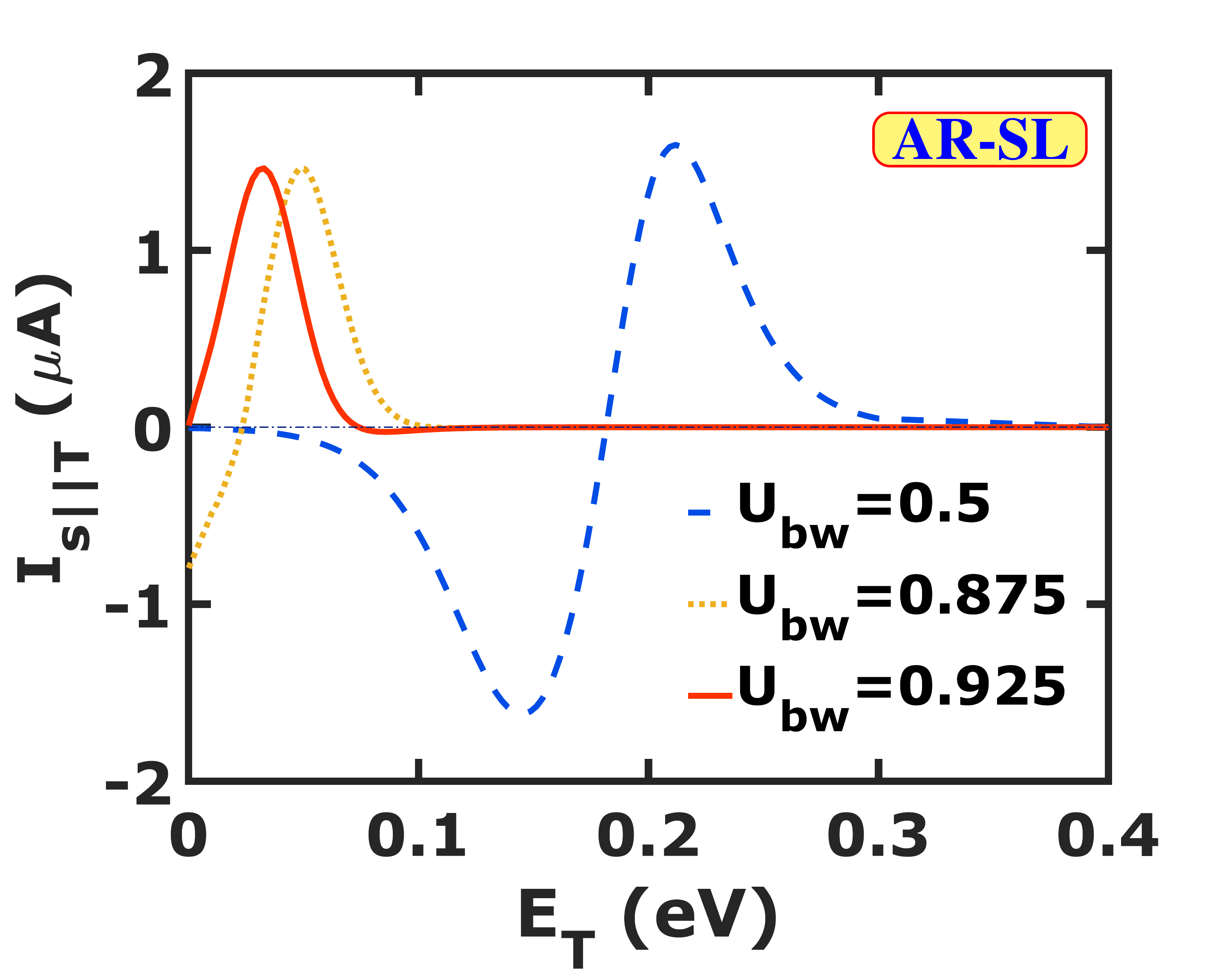}\label{ARSL_4B_IsPara_linMode_Ubw_90}}
	\quad
	\subfigure[]{\includegraphics[height=0.18\textwidth,width=0.225\textwidth]{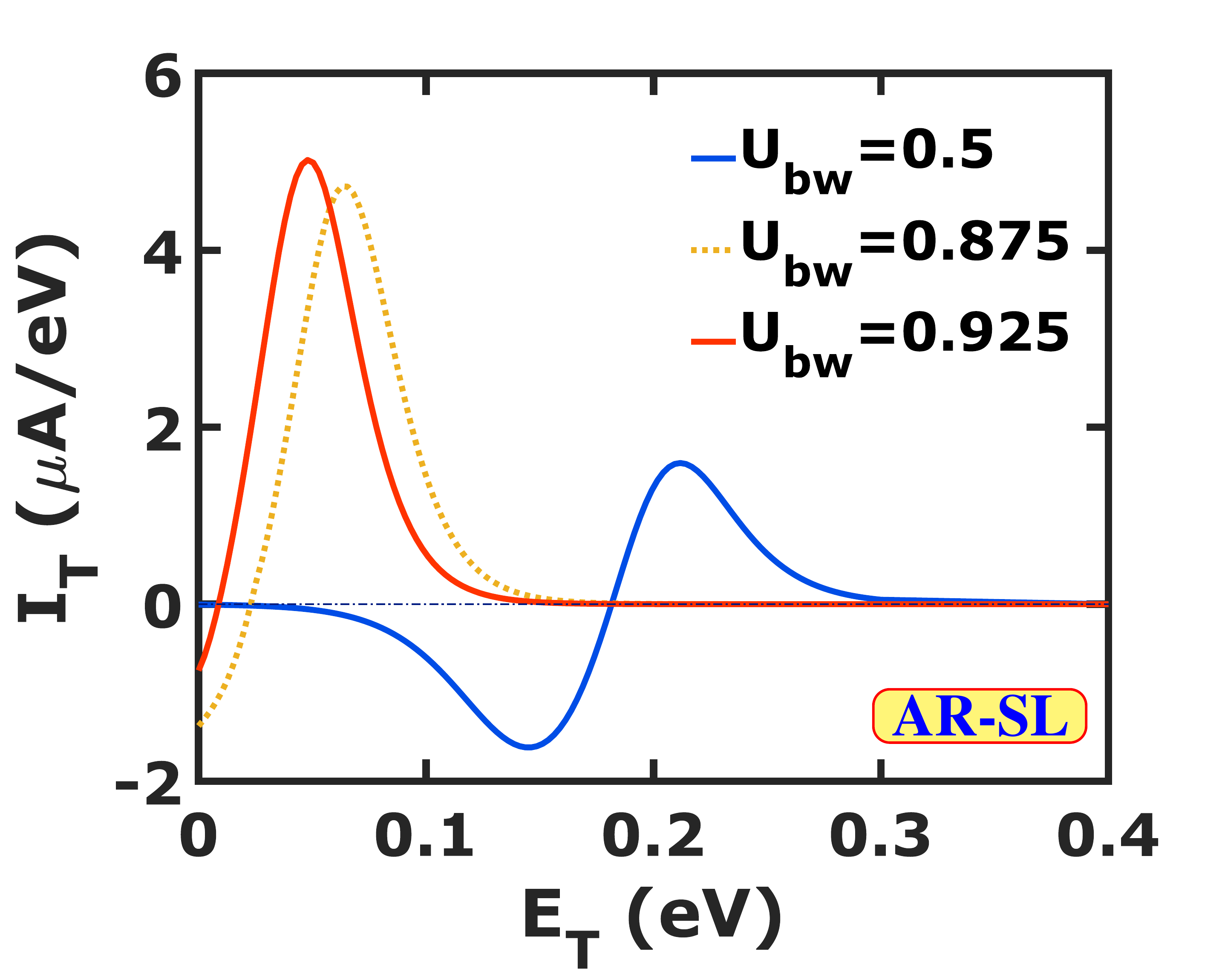}\label{ARSL_4B_Ic_linMode_Ubw_90}}
	
	\caption{(a) Parallel component of thermal spin transfer torque ($I_{s||}$) and (b) charge current ($I$) as a function of $U_{bw}$ for AR-SL in the perpendicular configuration of the fixed and free FMs, while keeping $V=0$ and $\Delta T=10 K$. Transverse mode profile for (c) parallel spin current and (d) charge current in a 4-barrier AR-SL.}
	\label{ARSL_4B_I_Upw_90}
\end{figure}

Now, we enable the AR-region in the same device configuration (as shown in Fig.~\ref{Device}(b)), with similar simulation setups and varying other parameters. It is quite interesting to note that at the same $U_{bw}=~0.925~eV$, a huge increment in the $I_{s||}$ as well as in the charge current $I$ as shown in Figs.~\ref{ARSL_4B_IsPara_Ubw_90} and \ref{ARSL_4B_Ic_Ubw_90}, respectively. This can be understood by comparing the transmission spectra plotted in Fig.~\ref{4B_TM_f1-f2_EP}, where the AR-SL configuration features a `boxcar' transmission that maximizes the transmissivity \cite{Whitney2014}. The boxcar feature only promotes the transmission in one direction as an analogue to the optical reflector. Furthermore, the transverse mode currents, shown in Figs.~\ref{ARSL_4B_IsPara_linMode_Ubw_90} and \ref{ARSL_4B_Ic_linMode_Ubw_90} gets amplified after using AR configuration.

\begin{figure}
	\subfigure[]{\includegraphics[height=0.18\textwidth,width=0.225\textwidth]{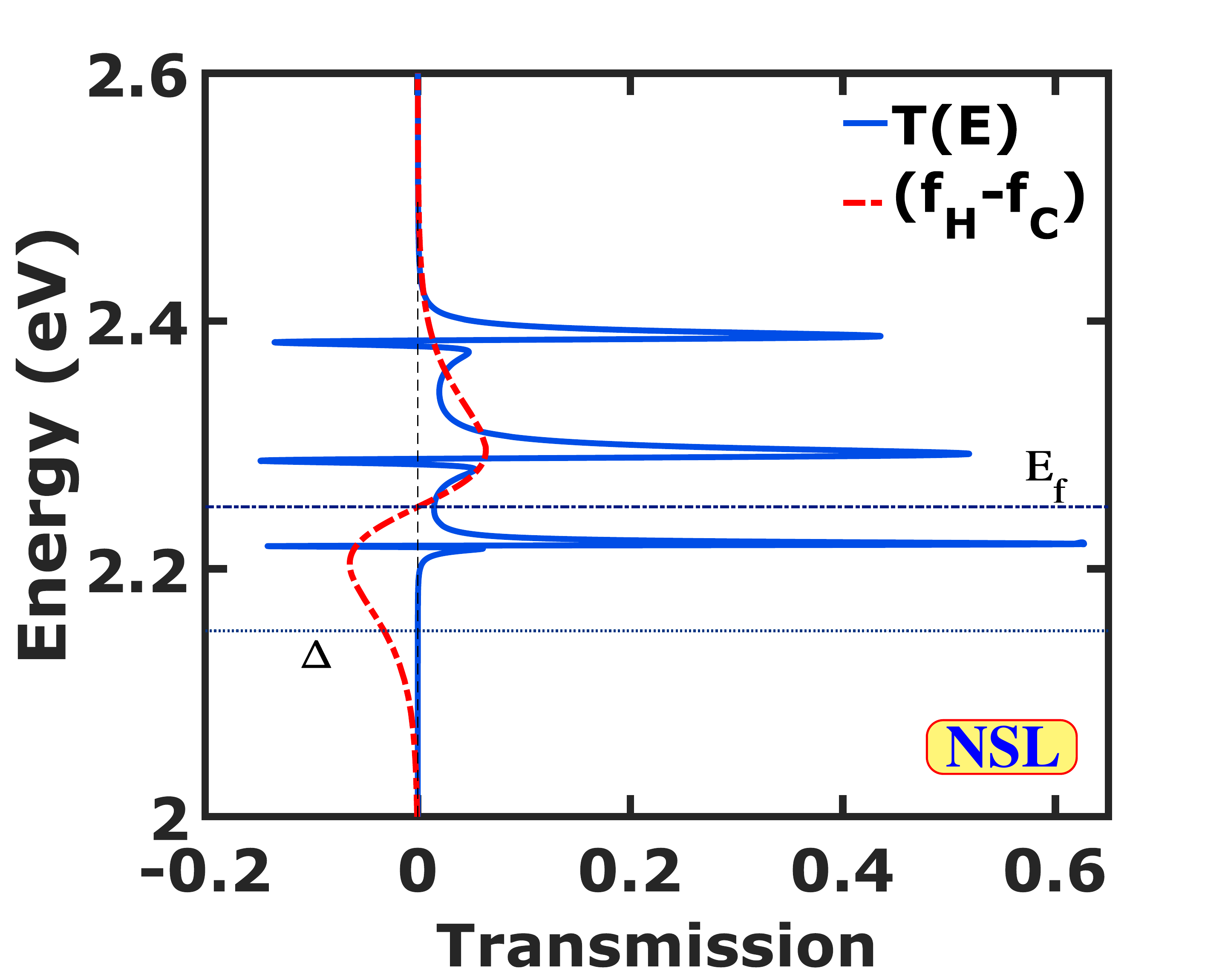}\label{NSL_4B_SpinTM_Fermi_Ubw0925_90}}
	\quad
	\subfigure[]{\includegraphics[height=0.18\textwidth,width=0.225\textwidth]{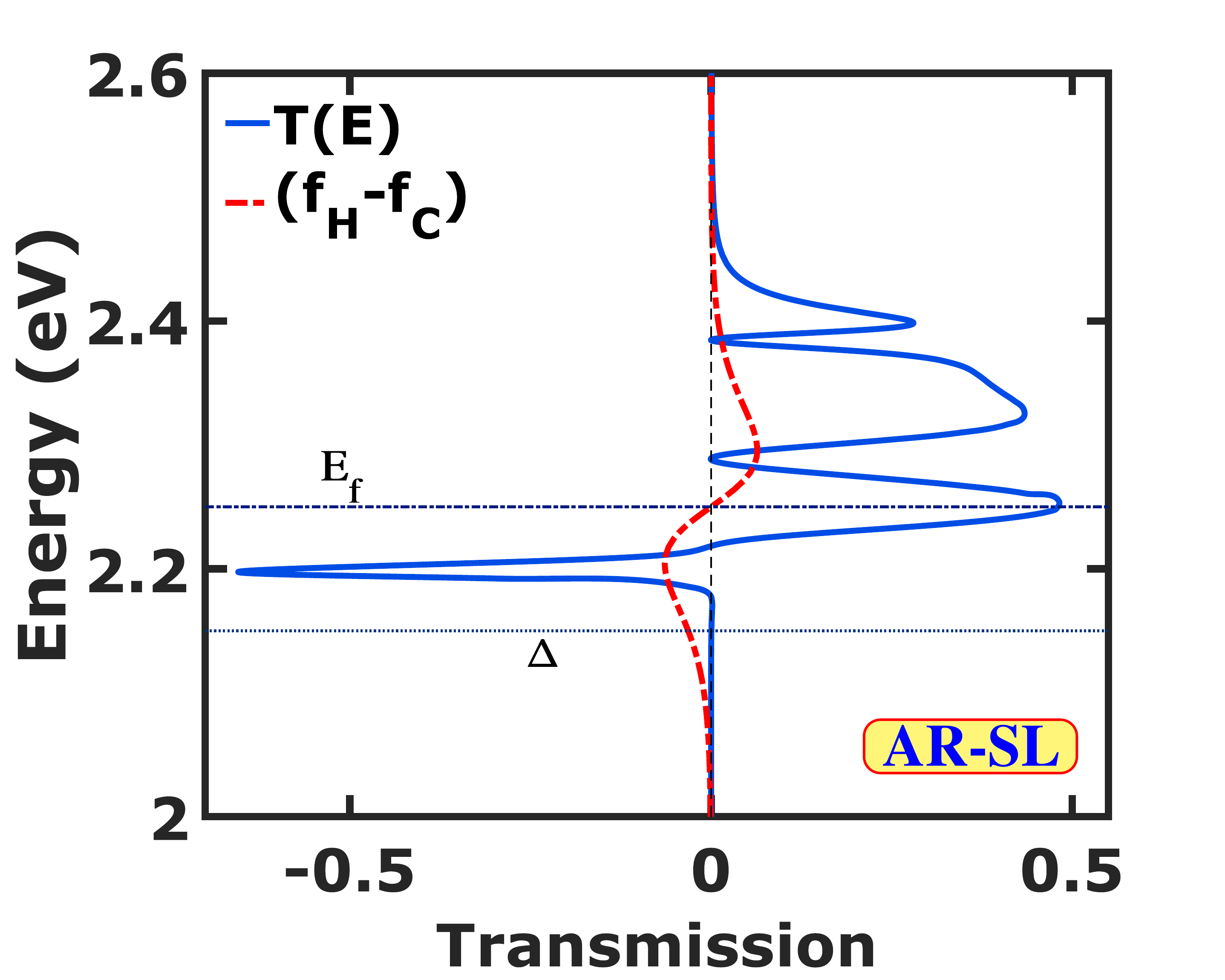}\label{_4B_SpinTM_Fermi_Ubw0925_90}}
	
	\caption{Color online: (a) Parallel component of spin resolved transmission coefficient as a function of energy $T(E)_{||}$ through a 4-barrier NSL (solid blue curve) for the first transverse mode with the Fermi function difference $(f_H-f_C)$ (red dash-dot line) at a temperature difference of 10 K between the contacts, when their magnetizations are perpendicular to each other. The labels $E_f$ and $\Delta$ indicate the Fermi level and exchange splitting respectively. Similarly (b), $T(E)_{||}$ coefficient as a function of energy $E$ through a 4-barrier AR-SL (solid blue curve) with Fermi function difference (red dash-dot line) at a temperature difference of 10 K between the contacts. Plots are zoomed in only for the first miniband.}
	\label{4B_SpinTM_f1-f2_90}
\end{figure}

In this analysis, we note that the value of charge current is larger than that of the spin current, that attributes to the transmission feature of the parallel component of spin current, plotted in Fig.~\ref{4B_SpinTM_f1-f2_90}. This Slonczewski (or parallel) term of spin current can also be expressed as 
\begin{equation}
I_{s,||} = T(E)_{||} . (f_H - f_C),
\end{equation}
where $T(E)_{||}$ is termed as the Slonczewski spin current transmission, given by
\begin{equation}
T(E)_{||} = \frac{1}{2} Trace[(\Gamma_{H}G\Gamma_{C}G^{\dagger} + G\Gamma_{C}G^{\dagger}\Gamma_{H}) S_M]
\end{equation}
where, $S_M= I\cross \sigma.\hat{M}$. The detailed derivation of non-equilibrium Slonczewski spin current and transmission can be found in supplementary notes of \cite{Abhishek2018}. Thus, it is quite interesting and unusual to have negative spin transmission as well, however, the transmission of charge transport is always positive.

\subsection{SL structural variation}
From the above analysis, the material parameters of the SL-MTJ nanowire are decided. We now fix the magnitude of $U_{bw}=~0.925~eV$ and $\Delta T=~10~K$ for both the SL configurations, for further analysis. Here, we examine the effect of well width variation, while, the barrier thickness is kept at $L_b=1.2~nm$. It is well known fact that the currents through a thicker barrier get diminished, thus we strike off the barrier thickness variation in order to optimize the SL structure.

\begin{figure}
	\subfigure[]{\includegraphics[height=0.18\textwidth,width=0.225\textwidth]{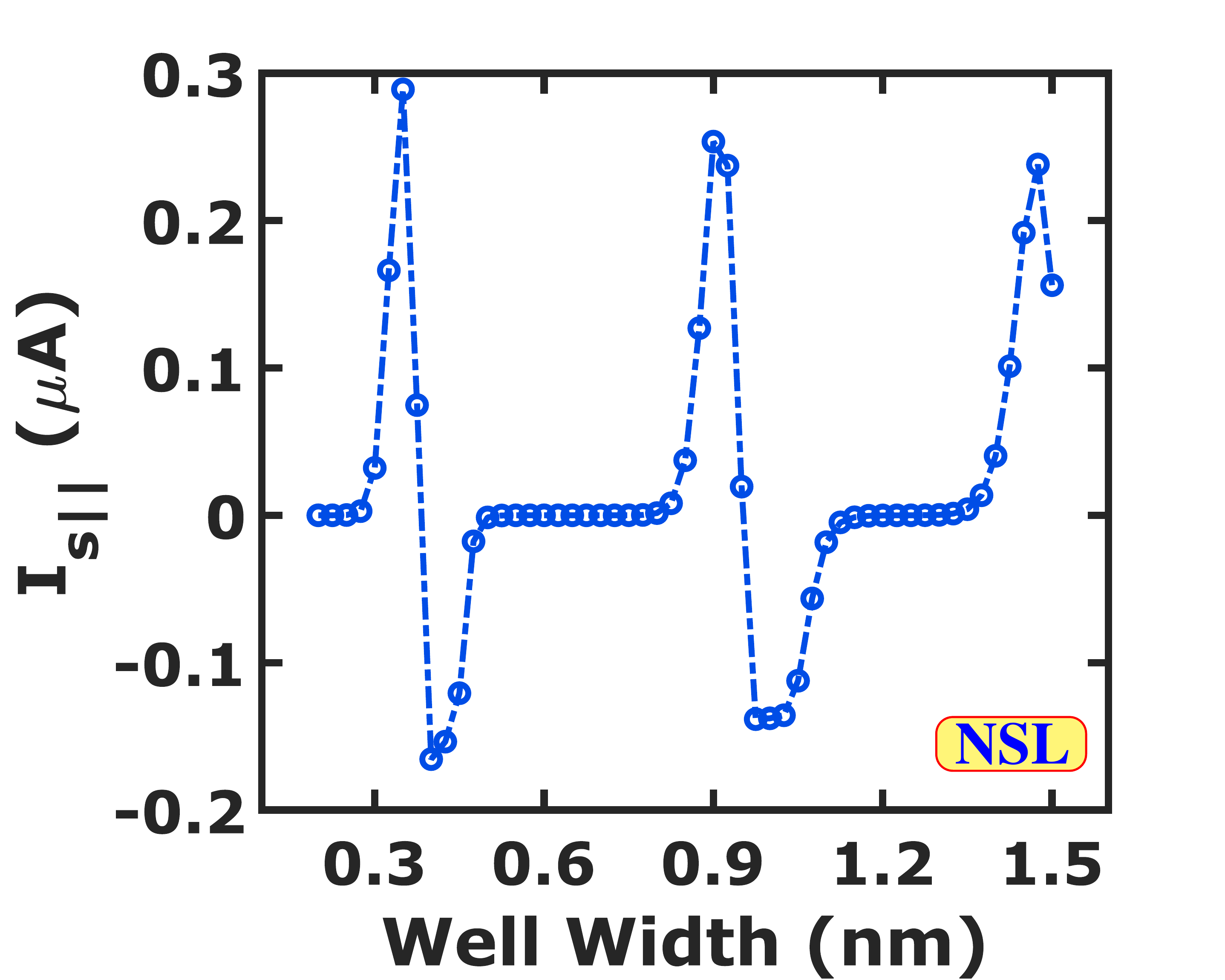}\label{NSL_4B_IsPara_LW_90}}
	\quad
	\subfigure[]{\includegraphics[height=0.18\textwidth,width=0.225\textwidth]{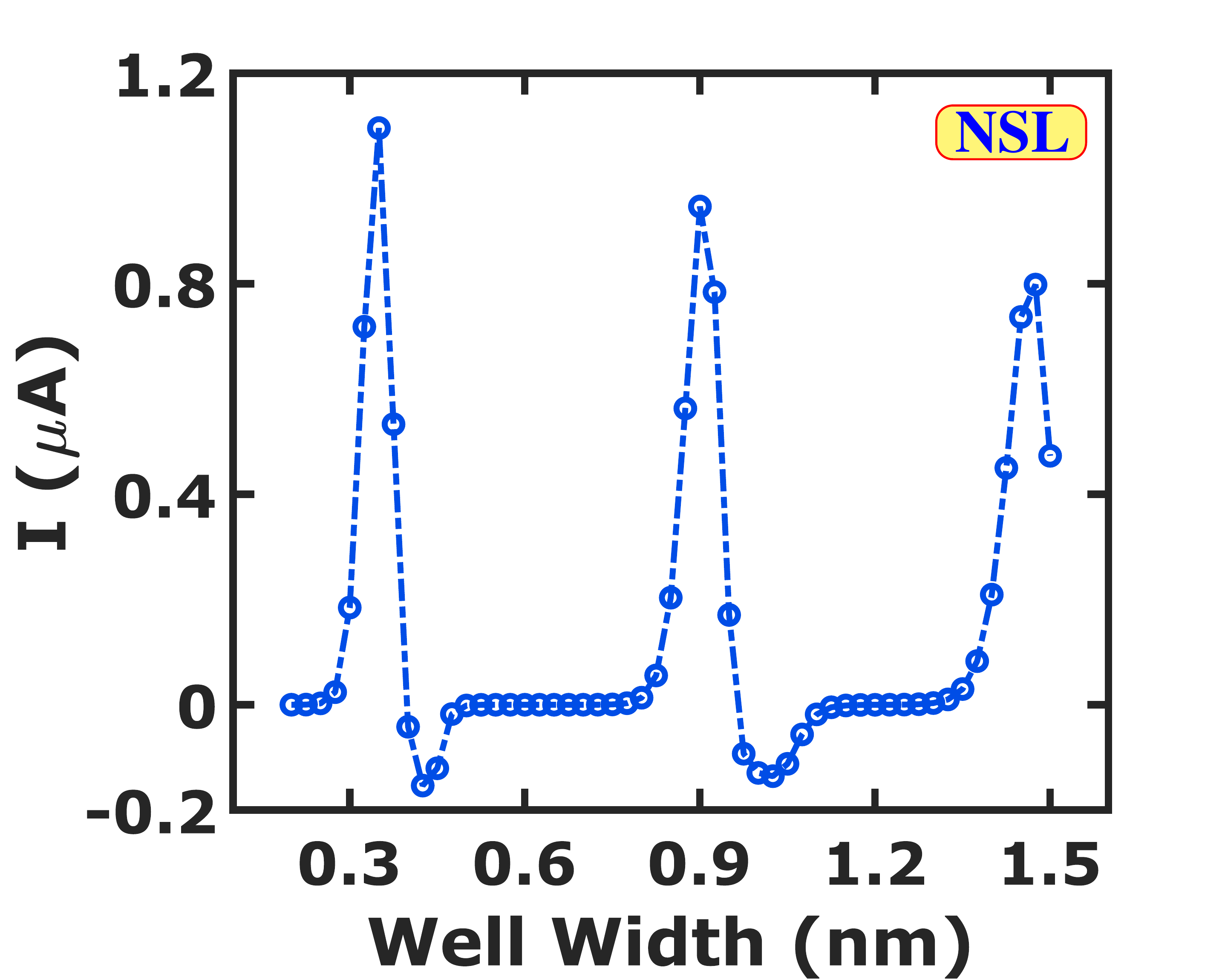}\label{NSL_4B_Ic_LW_90}}
	\quad
	\subfigure[]{\includegraphics[height=0.18\textwidth,width=0.225\textwidth]{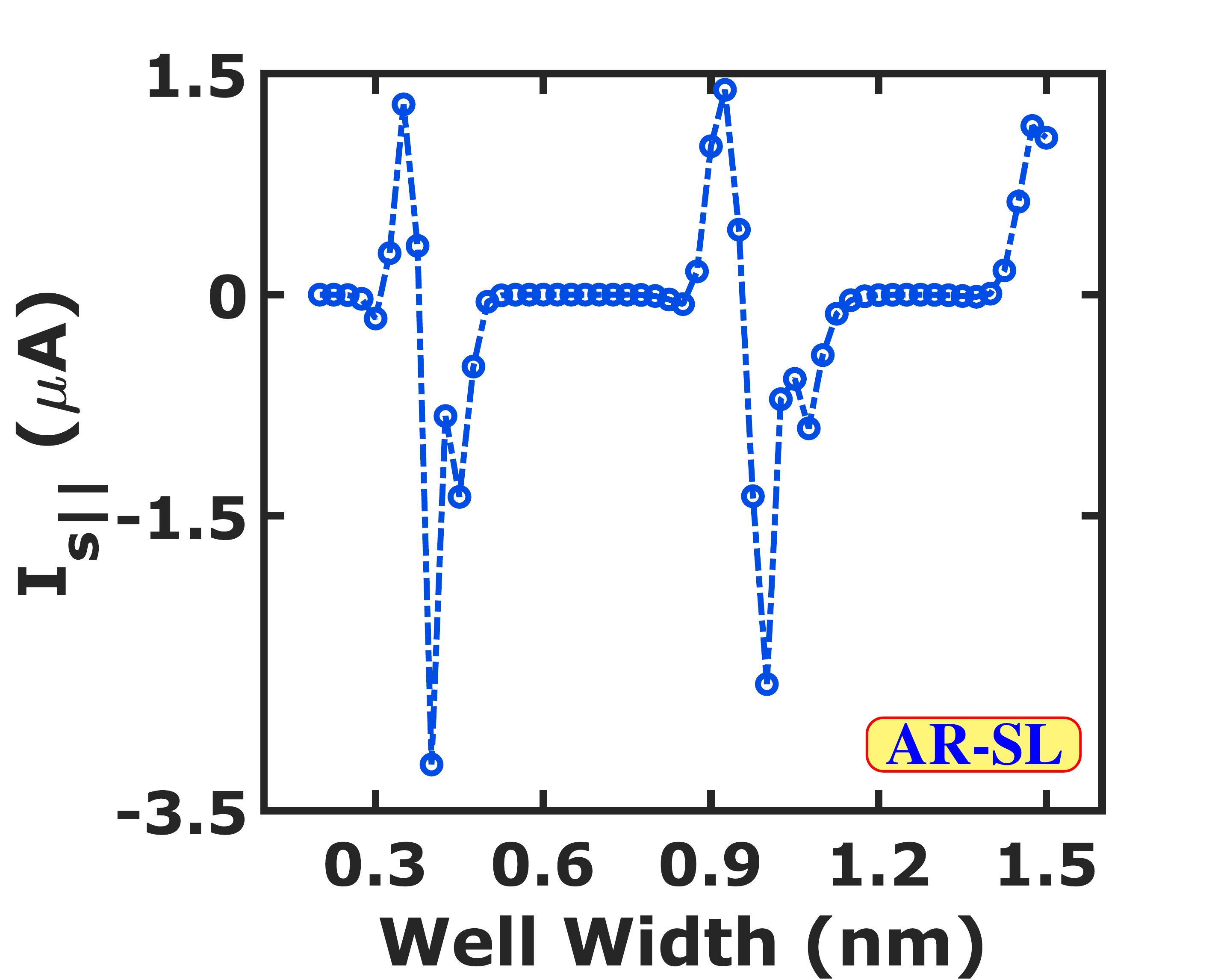}\label{ARSL_4B_IsPara_LW_90}}
	\quad
	\subfigure[]{\includegraphics[height=0.18\textwidth,width=0.225\textwidth]{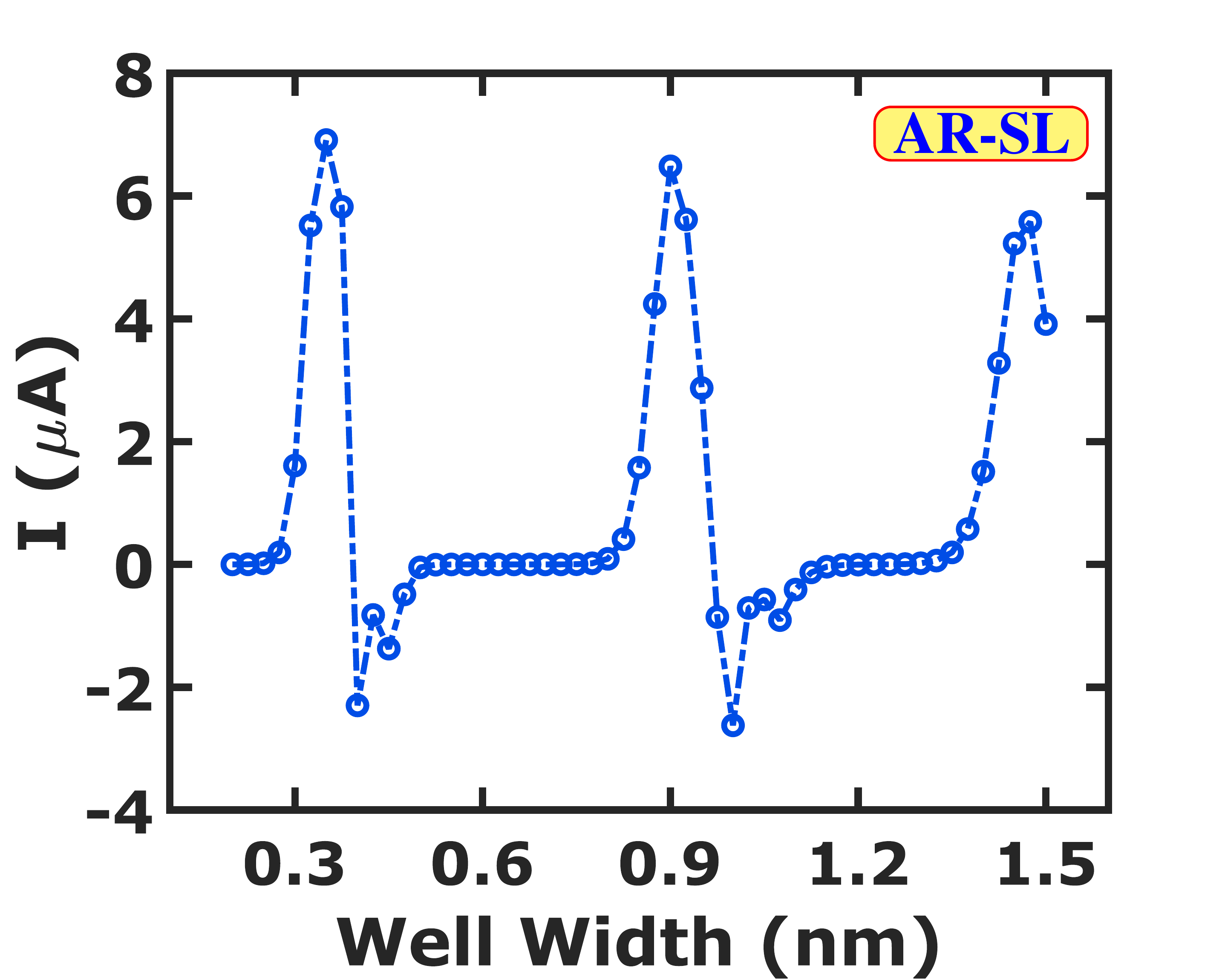}\label{ARSL_4B_Ic_LW_90}}
	
	\caption{(a) Parallel component of thermal spin transfer torque ($I_{s||}$) and (b) charge current ($I$) as a function of well width ($L_{w}$) for NSL in the perpendicular configuration of the fixed and free FMs, while keeping $V=0$ and $\Delta T=10 K$. Likewise, (c) TSTT and (d) charge current as function of $L_w$ for AR-SL.}
	\label{4B_I_LW_90}
\end{figure}

We plot the variation of $I_{s||}$ and $I$ as a function of well width, in Figs.~\ref{4B_I_LW_90}, for both SL configurations. The overall sinusoidal feature of currents depicts that, as the well width increases, the number of quantum states in the well region increases in the vicinity of $E_f$. The value of $I_{s||}$ oscillates with increasing well width in both the configurations as shown in Figs.~\ref{NSL_4B_IsPara_LW_90} \& \ref{ARSL_4B_IsPara_LW_90}, with a larger value in the case of AR-SL. Likewise, the charge current $I$ oscillates with increasing well width in both cases. Here, we note that the negative value of $I_{s||}$ is more than that of the charge current $I$, and this due to the negative transmission feature of the parallel component of spin current. Eventually, we take $L_w=0.35~nm$ for the above results that are plotted in Figs.~\ref{4B_TM_f1-f2_EP}, \ref{NSL_4B_I_Upw_90}, \ref{ARSL_4B_I_Upw_90}, and \ref{4B_SpinTM_f1-f2_90}.

\subsection{Comparision between various MTJ structures}
Now we compare the results with the size of SL, that depends on the number of barriers used in the device central region as depicted in Fig.~\ref{Device}. This could guide us to optimize the device length in the context to maximize our requirements. In order to carry out this comparison, we use the same material \& structural parameters that have been used so far in the linear regime of thermal bias only.

\begin{figure}
	\subfigure[]{\includegraphics[height=0.18\textwidth,width=0.225\textwidth]{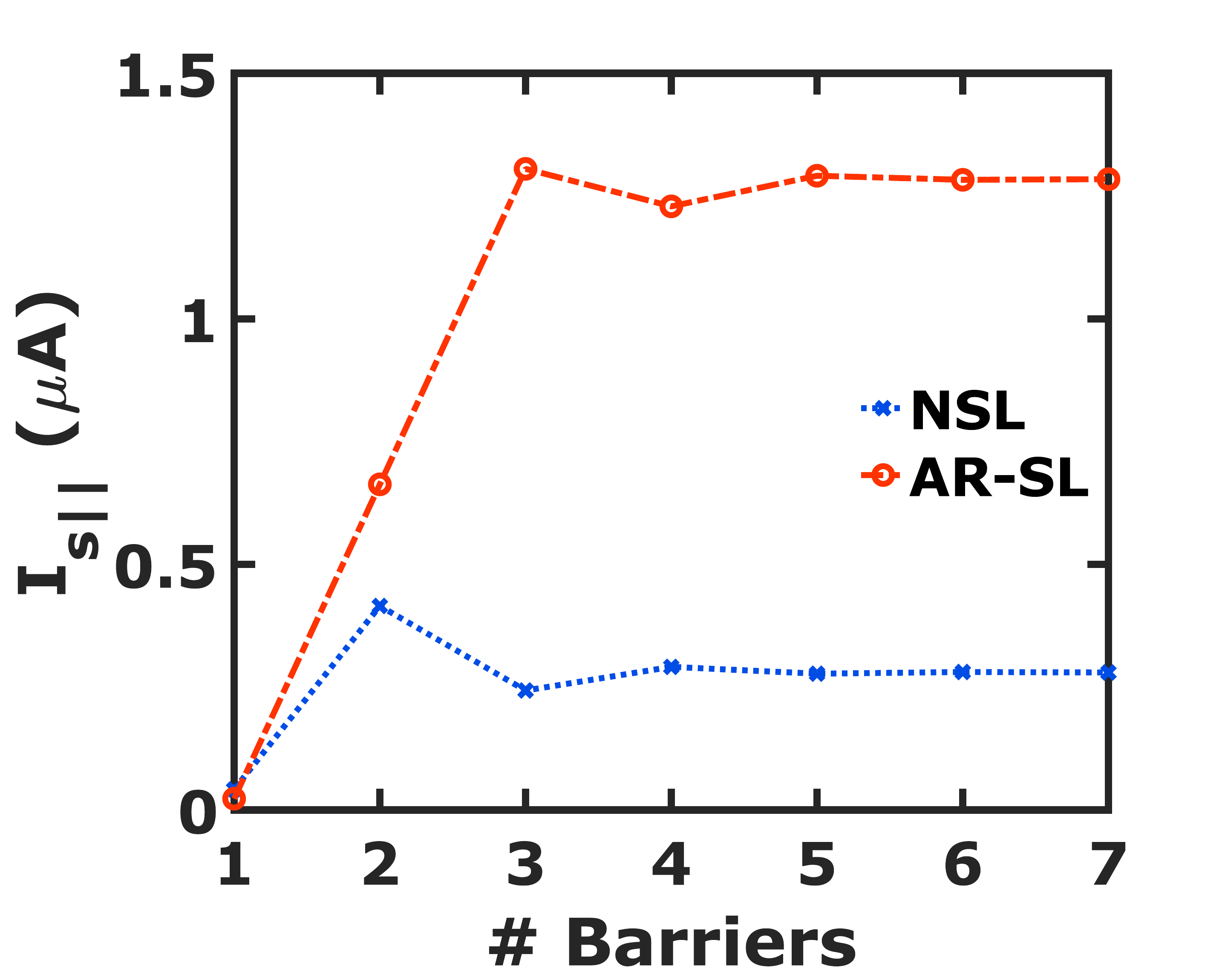}\label{Barrier_IsPara_90}}
	\quad
	\subfigure[]{\includegraphics[height=0.18\textwidth,width=0.225\textwidth]{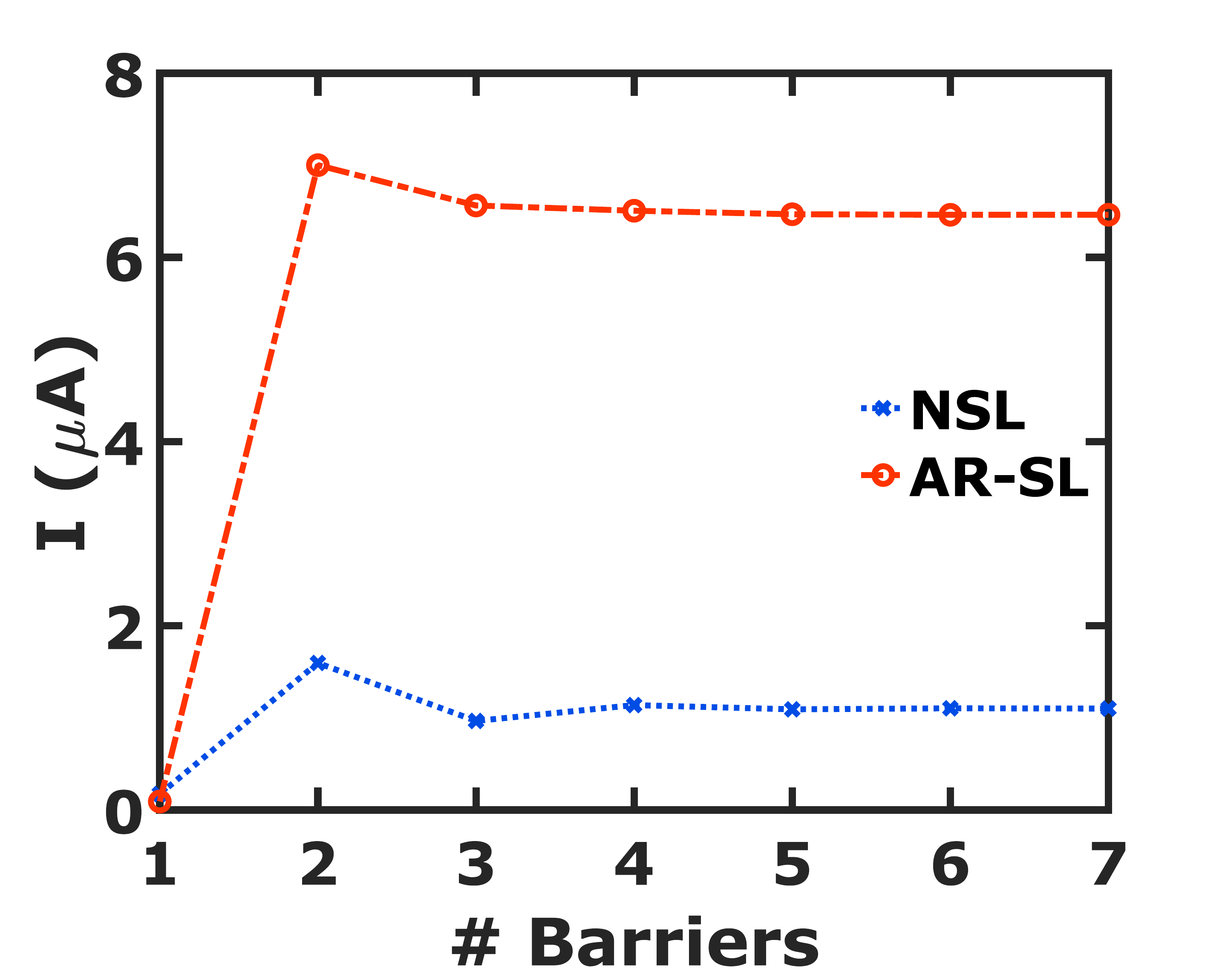}\label{Barrier_Ic_90}}
	
	\caption{(a) Thermal spin torque and (b) charge current as a function of number of barriers in NSL and AR-SL, while keeping $V=0$ and $\Delta T=10 K$ with above extracted parameters.}
	\label{Barrier_I_90}
\end{figure}

We show in Fig.~\ref{Barrier_I_90} the TSTT and charge current variation with the number of barriers in the SL configurations. Here, we may refer MTJ family as the variety of SL configurations with different number of barriers in the structure \cite{Pankaj2018}, sandwiched between magnetic contacts. In our study, we consider only the normal (NSL) and the anti-reflective superlattice (AR-SL) configurations. From Fig.~\ref{Barrier_IsPara_90}, we note that the TSTT first increases up to 3 barriers and then almost saturates afterward in both the configurations, but the magnitude of TSTT in AR-SL is almost five times greater than that of NSL. However, the flow of charge current settles down approximately after 2 barriers, but manages the same enhancement in the current magnitudes. In the SL structures, as the number of barriers increase (accordingly number of well also increases), the number of transmission peaks (equal to the quantum-well in the structure) within the band-pass spectra also increases. Consequently, the transmissivity gets reduced beyond a certain number of barriers, and so thus the currents.

\subsection{Temperature variation}
\begin{figure}
	\subfigure[]{\includegraphics[height=0.18\textwidth,width=0.225\textwidth]{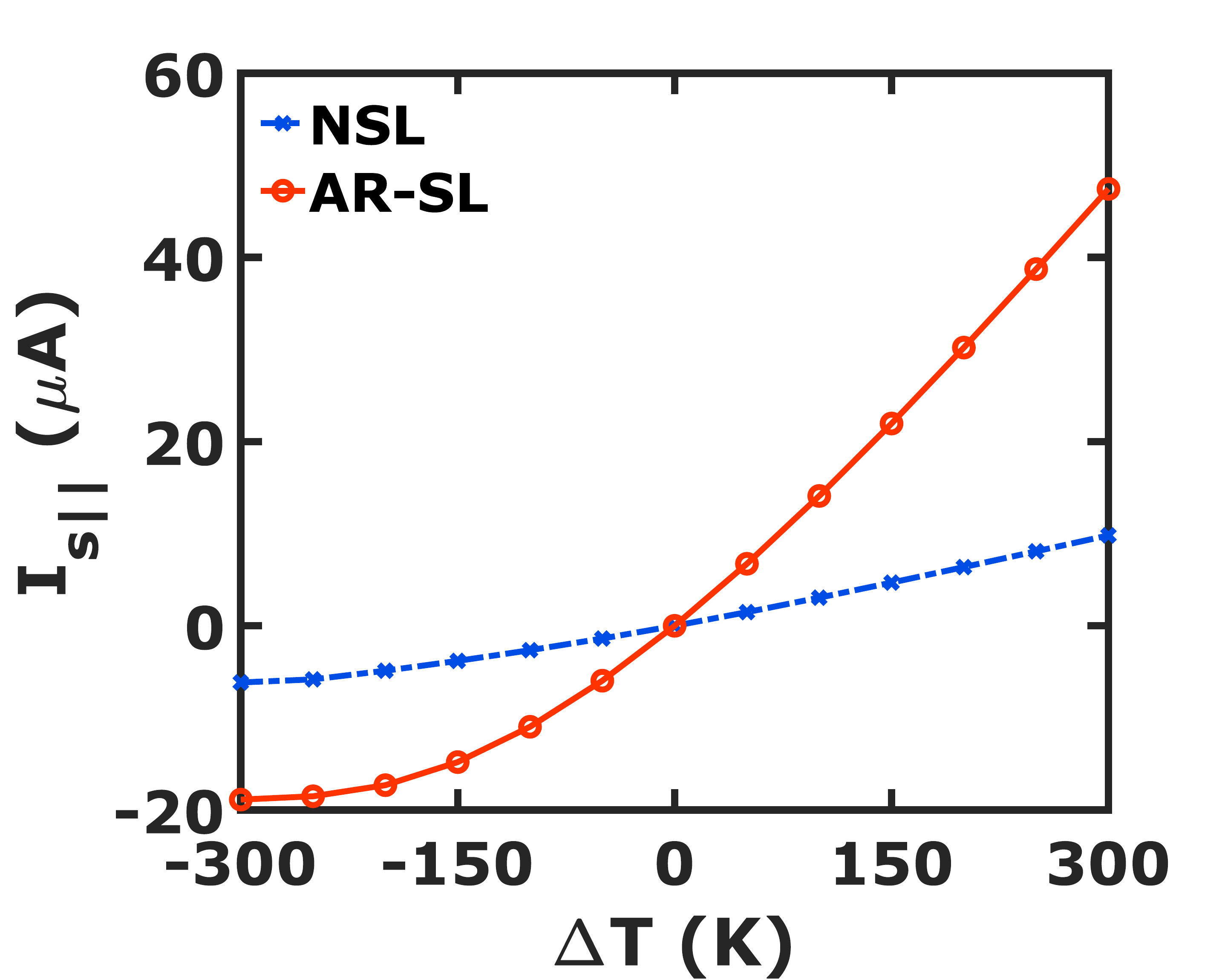}\label{NSL_ARSL_4B_IsPara_deltaT_90}}
	\quad
	\subfigure[]{\includegraphics[height=0.18\textwidth,width=0.225\textwidth]{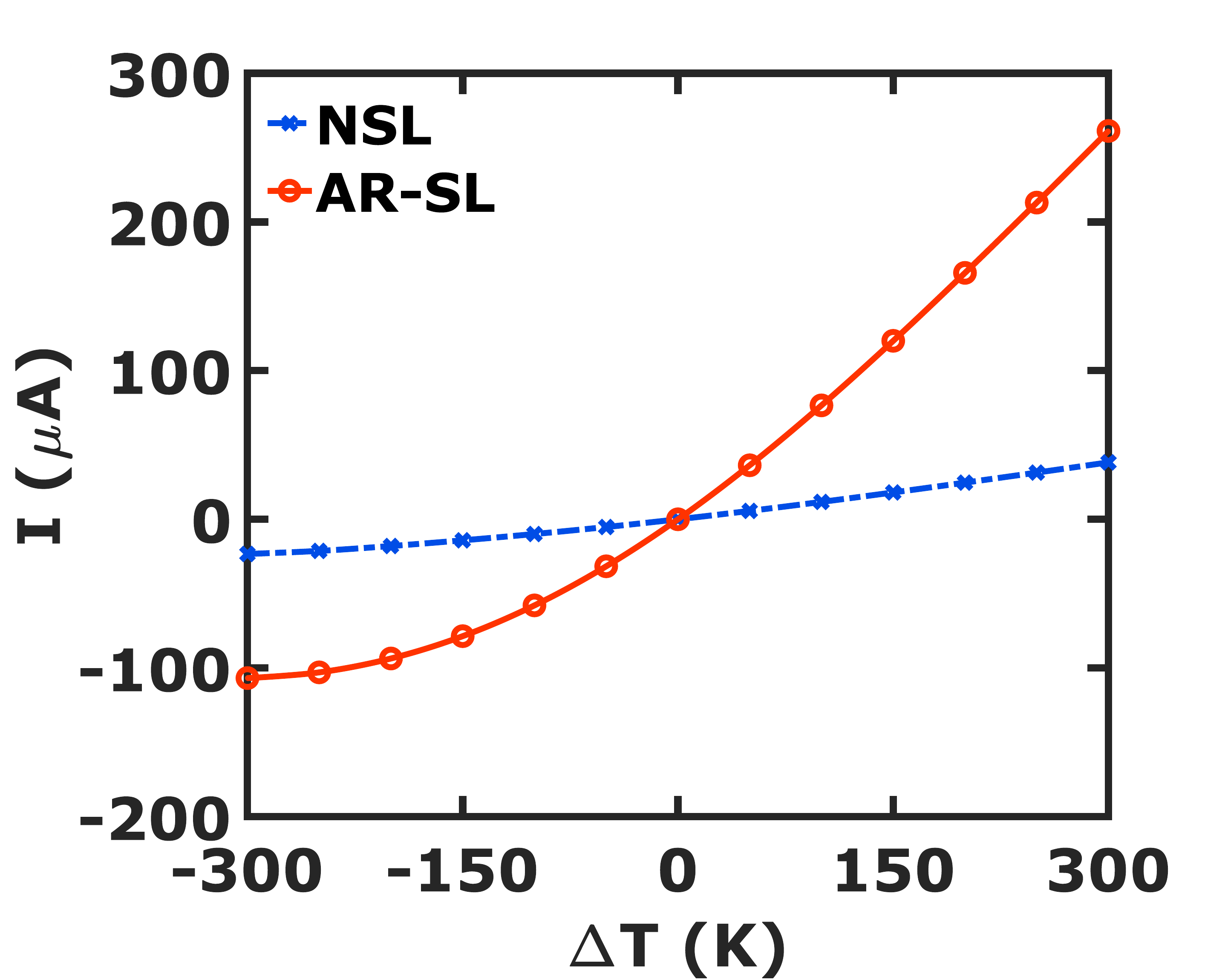}\label{NSL_ARSL_4B_Ic_deltaT_90}}
	
	\caption{(a) TSTT for the NSL and AR-SL configuration as a function of the temperature difference $\Delta T$ between the fixed and the free FM contacts. (b) Similarly, charge current $I$ as a function of $\Delta T$ in the perpendicular configuration of the free and fixed FMs.}
	\label{4B_I_deltaT_90}
\end{figure}

So far, we have fixed the temperature difference between the contacts to meet the linear response theory. This theory represents the
linear response of the system to the external perturbation, in order to analyze the results. In practice, even a very small thermal gradient is enough to produce this requirement \cite{Arnab2018}.

In order to know the effect of wide variation of temperature difference on the device, the TSTT and charge current as a function of $\Delta T$ is plotted in Fig.~\ref{4B_I_deltaT_90}. In the plots, negative $\Delta T$ means that the contact attached with free FM is kept at a higher temperature than the fixed FM contact and vice versa. In the case of NSL, on comparing between Figs.~\ref{NSL_ARSL_4B_IsPara_deltaT_90} and \ref{NSL_ARSL_4B_Ic_deltaT_90}, we notice that the magnitude of $I_{s||}$ and charge current ($I$) does not depend on the sign of $\Delta T$. The change in sign of $\Delta T$ results only to the change of direction of the torque exerted on the free FM layer. We note from Fig.~\ref{4B_I_deltaT_90} that there is a non-linear behavior of $I_{s||}$ and $I$ at higher temperature difference in AR-SL configuration, that restricts us to analyze in the linear regime of operation. 

In this analysis, the negative spin current with negative $\Delta T$ depicts the reversal of torque in the direction on the free FM. This effect is quite in contrast to the condition when only a finite voltage bias is applied across the contacts \cite{Abhishek2018}. In such a situation, not only the direction but also the magnitude of the torque can depend on the sign of the bias voltage. Customarily, it might be argued that the negative currents obtained so far support the p-type conduction.


\section{Conclusion}
In conclusion, the thermal spin transfer torque in the multi-barrier normal and anti-reflective superlattice structures is studied, using spin-resolved NEGF formalism framework. We discuss the view on capturing full swing of currents using nonmagnetic metals in the well region to eliminate the negative contributions due to the asymmetry in Fermi difference. We have optimized all the essential material and structural parameters to get a larger boost in the TSTT. Interestingly, there are a huge enhancement, approximately five times in magnitude of TSTT as well as in the charge current for the AR-SL than that of the NSL. This routes the uses of a ``boxcar" type \cite{Whitney2014} bandpass filtering approach enabled through the AR region. Although, the thermal STT is smaller than the voltage driven STT, in general, this investigation establishes a more efficient way to switch the magnetization of the free FM without any electrical losses. This temperature gradient may be achieved with short laser pulses \cite{Choi2015} in practice. The role of the perpendicular component of TSTT, however, is still unclear and needs more discussion. With the existing advanced thin-film growth technology, the discussed superlattice configurations can be achieved, and such an optimized TSTT functionality can be realized.\\
{\it{Acknowledgements:}} This work is an outcome of
the Research and Development work undertaken in the project
under the Visvesvaraya Ph.D. Scheme of Ministry of Electronics and Information Technology, Government of India, being
implemented by Digital India Corporation (formerly Media
Lab Asia). This work was also supported by the Science and
Engineering Research Board (SERB) of the Government of
India under Grant No. EMR/2017/002853.

\bibliographystyle{IEEEtran}
\bibliography{Reference}

\end{document}